\documentclass[journal=nalefd,manuscript=letter,layout=twocolumn]{achemso}

\usepackage[version=3]{mhchem} 
\usepackage{graphicx}
\usepackage{dcolumn}
\usepackage{bm}
\usepackage[mathlines]{lineno}
\usepackage{amssymb}
\usepackage{amsmath}
\usepackage{natbib}
\usepackage{soul,color}

\title{Unusually stable helical coil allotrope of phosphorus}

\author{Dan~Liu}
\affiliation{Physics and Astronomy Department,
             Michigan State University,
             East Lansing, Michigan 48824, USA}

\author{Jie~Guan}
\affiliation{Physics and Astronomy Department,
             Michigan State University,
             East Lansing, Michigan 48824, USA}

\author{Jingwei~Jiang}
\affiliation[Permanent address:  ]{Department of Physics,
             Peking University,
             Beijing, 100871, China}
\affiliation{Physics and Astronomy Department,
             Michigan State University,
             East Lansing, Michigan 48824, USA}

\author{David Tom\'{a}nek}
\affiliation{Physics and Astronomy Department,
             Michigan State University,
             East Lansing, Michigan 48824, USA}
\email
            {tomanek@pa.msu.edu}%
\affiliation{Physics and Astronomy Department,
             Michigan State University,
             East Lansing, Michigan 48824, USA}

\date{\today} 

\abbreviations{DFT, PBE, LDA, DOS}


\begin{document}


\begin{abstract}
We have identified an unusually stable helical coil allotrope of
phosphorus.
Our {\em ab initio} Density Functional Theory calculations
indicate that the uncoiled, isolated straight 1D chain is equally
stable as a monolayer of black phosphorus dubbed phosphorene. The
coiling tendency and the attraction between adjacent coil segments
add an extra stabilization energy of ${\approx}12$~meV/atom to the
coil allotrope, similar in value to the ${\approx}16$~meV/atom
inter-layer attraction in bulk black phosphorus. Thus, the helical
coil structure is essentially as stable as black phosphorus, the
most stable phosphorus allotrope known to date. With an optimum
radius of 2.4~nm, the helical coil of phosphorus may fit well and
even form inside wide carbon nanotubes.
\end{abstract}



\begin{figure*}[t]
\includegraphics[width=1.6\columnwidth]{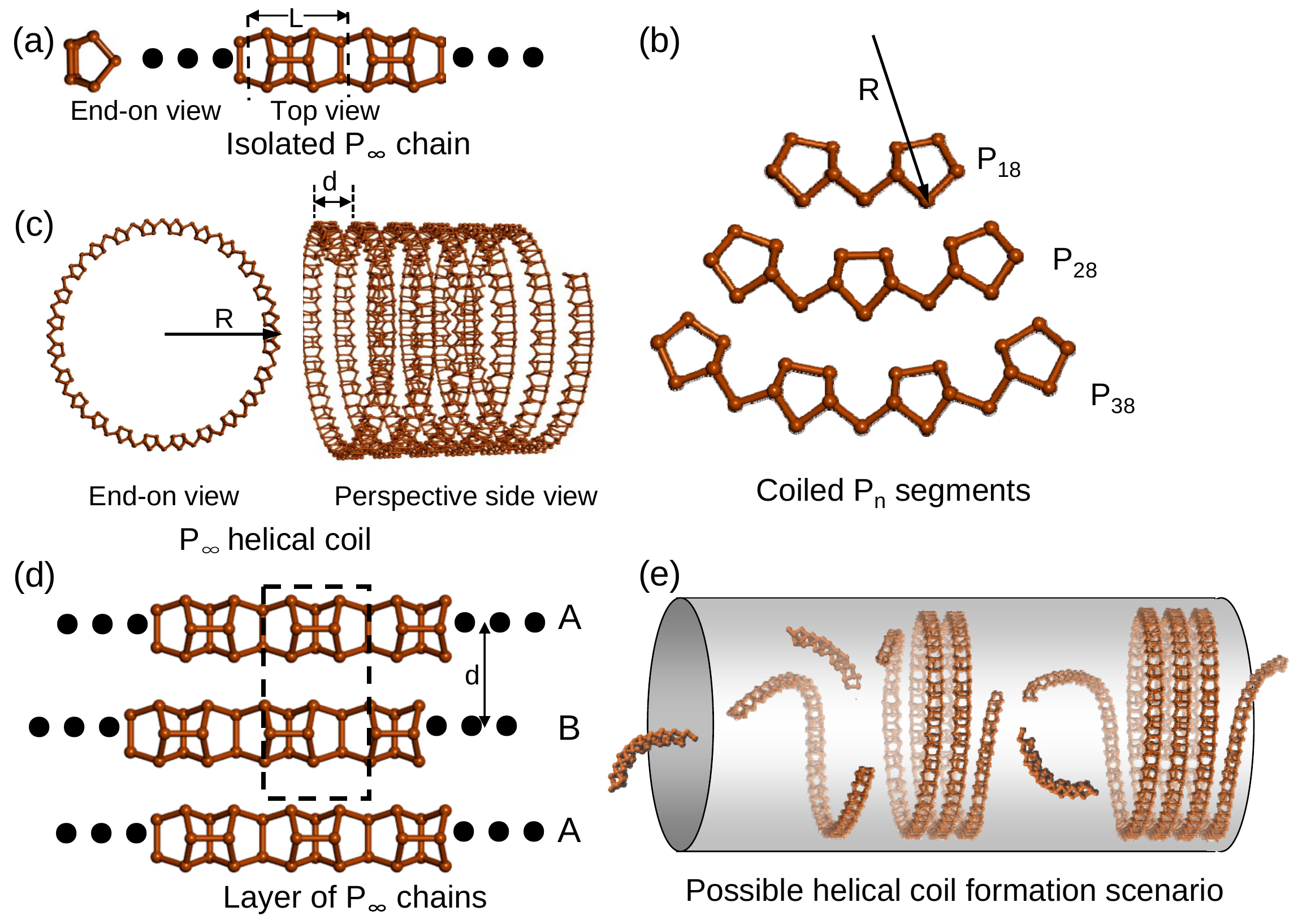}
\caption{(Color online) (a) Optimum structure of an isolated,
straight 1D P$_\infty$ chain with the P$_{10}$ unit cell of length
$L$. (b) Optimized P$_{18}$, P$_{28}$ and P$_{38}$ segments of the
isolated chain, indicating the tendency to form rings with a
radius $R{\approx}2.4$~nm. (c) The optimum structure of a single
coil. (d) Optimum structure of a 2D assembly of P$_\infty$ chains
separated by the distance $d$. (e) Possible scenario for the
formation of helical coils by connecting finite-length chain
segments inside a cylindrical cavity. The unit cells are indicated
by dashed lines in (a) and (d). \label{fig1}}
\end{figure*}

Elemental phosphorus has been known for its unusual properties
since its isolation as the white phosphorus
allotrope~\cite{Brandt1669}, a P$_4$-based molecular solid, in
1669. Other observed allotropes include violet
phosphorus~\cite{{violetp},{Thurn69}}, also known as Hittorf's
metallic phosphorus~\cite{{Hittorf1865},{Ruck2005}}, rather common
fibrous red phosphorus~\cite{{redp-blackp35},{Ruck2005}} with an
amorphous structure, and layered black phosphorus~\cite{blackP14},
known as the most stable crystalline allotrope. Other bulk
allotropes, including blue phosphorene, have been
predicted~\cite{DT230} and subsequently
synthesized~\cite{Zhang2016NL}. Other structures of elemental
phosphorus, which have been studied, include P$_n$
clusters~\cite{{Haser92},{SeifertJCP92},{SeifertZPD93}} and
atomically thin P helices, which have been identified as
constituents~\cite{Pfister16} in the complex structure of SnIP. It
thus appears quite possible that still more allotropes may be
synthesized in the future.


We report here theoretical results that identify an unusually
stable helical coil allotrope of phosphorus. Our {\em ab initio}
Density Functional Theory calculations indicate that the uncoiled,
isolated straight 1D chain is equally stable as a monolayer of
black phosphorus dubbed phosphorene. The coiling tendency and the
attraction between adjacent coil segments add an extra
stabilization energy of ${\approx}12$~meV/atom to the coil
allotrope, similar in value to the ${\approx}16$~meV/atom
inter-layer attraction in bulk black phosphorus. Thus, the helical
coil structure is essentially as stable as black phosphorus, the
most stable phosphorus allotrope known to date. With an optimum
radius of 2.4~nm, the helical coil of phosphorus may fit well and
even form inside wide carbon nanotubes.

\section{Equilibrium structure and stability results}

The unusually stable structure of a P$_{10}$ cluster and its
suitability to link up to an infinite 1D chain was discovered
while developing and testing a Genetic Algorithm optimization
technique for phosphorus clusters based on a tight-binding
formalism~\cite{DLiuTB16}. The optimum structure of a P$_{10}$
unit cell in a straight 1D chain, which resembles a narrow tube
with a pentagonal cross-section, is shown in Fig.~\ref{fig1}(a).
We notice a structural similarity with fibrous
red~\cite{{redp-blackp35},{Ruck2005}} and violet
phosphorus~\cite{{Hittorf1865},{violetp},{Thurn69},{Ruck2005}}
structures, which also contain P$_{10}$ subunits in the
interlinked chains. The postulated chain structure is also similar
to P nanorods~\cite{{Pfitzner04},{Bachhuber14}} and P
tubes~\cite{Grotz15} observed in the AgP$_{15}$ compound. Our
DFT-PBE calculations indicate a binding energy
$E_{coh}=3.274$~eV/atom for the postulated P$_{10}$ structure with
respect to spin-polarized P atoms. This value is only negligibly
larger than that of a monolayer of black phosphorus, known as the
most stable phosphorus allotrope, with $E_{coh}=3.273$~eV/atom.

\begin{figure}[t!]
\includegraphics[width=0.7\columnwidth]{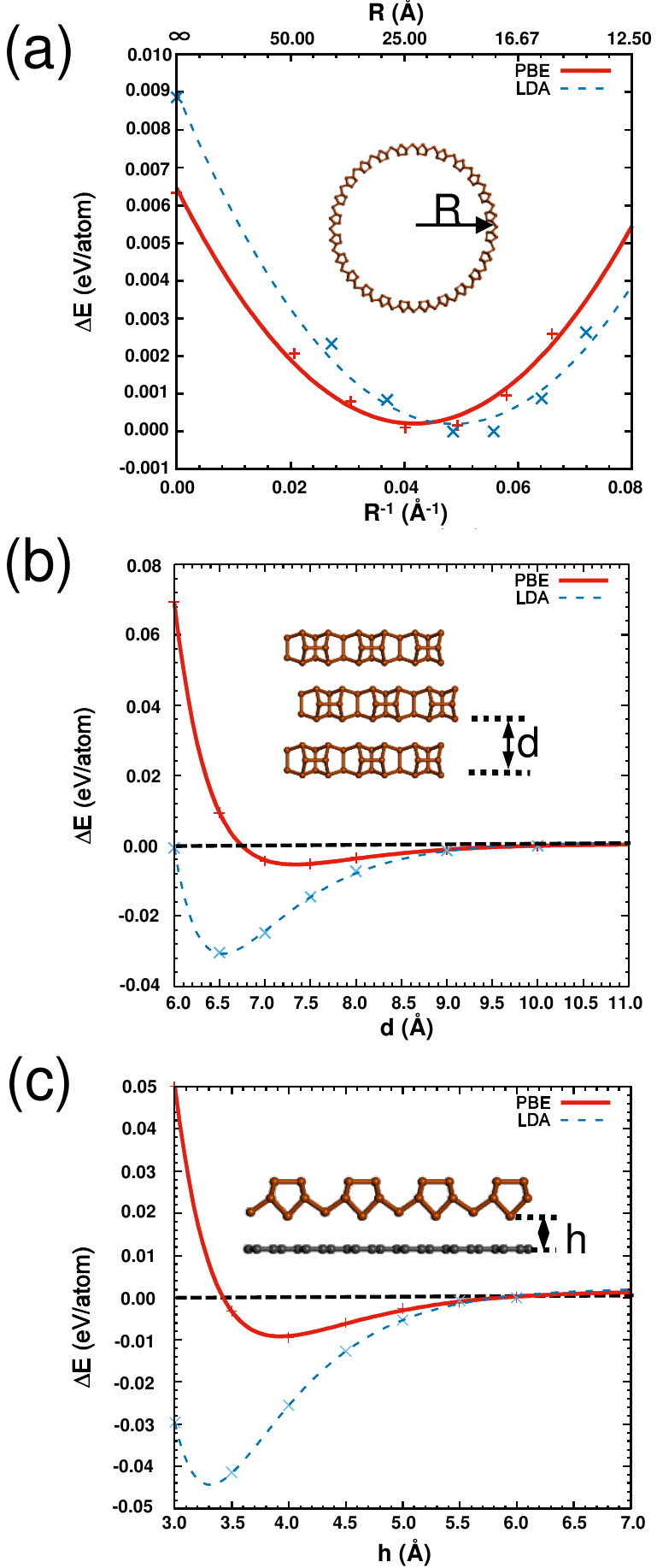}
\caption{(Color online) (a) Strain energy ${\Delta}E$ per atom as
a function of the radius $R$ of an isolated P coil. (b)
Inter-chain interaction energy ${\Delta}E$ per atom in a 2D
assembly of phosphorus chains, depicted in
Fig.~\protect\ref{fig1}(d), as a function of the inter-chain
distance $d$. (c) Interaction energy ${\Delta}E$ per phosphorus
atom between a P chain and a graphene monolayer as a function of
the adsorption height $h$. PBE results are shown by the solid red
lines, LDA results by the dashed blue lines. \label{fig2}}
\end{figure}

Finite chain segments, shown in Fig.~\ref{fig1}(b), display a
tendency to form coils with an average radius of $2.4$~nm. We find
this coiling, which had been identified
earlier~\cite{{Haser92},{Karttunen2007}}, to be associated with an
energy gain of $6$~meV/atom (PBE) and $9$~meV/atom (LDA). Assuming
that an ordered P$_\infty$ system may form by connecting finite
P$_n$ segments end-to-end, the resulting equilibrium structure
will be a helical coil, depicted in Fig.~\ref{fig1}(c). On a
per-atom basis, the elastic strain ${\Delta}E$ in the coil is
shown in Fig.~\ref{fig2}(a) as a function of radius $R$. The data
points are well represented by
\begin{equation}
{\Delta}E = \frac{1}{2} k %
\left( \frac{1}{R} - \frac{1}{R_{eq}}\right)^2 \,, %
\label{eq1}
\end{equation}
where $R_{eq}$ is the equilibrium radius. This expression
describes the local strain energy in a finite-length 1D beam
thought to be initially aligned with the $x-$direction and
deformed to a circular arc of radius $R$ in the $xz-$plane. The
local strain is $\sigma=d^2u_z/dx^2=1/R$. Should not a straight,
but rather a bent beam of radius $R_{eq}$ represent the
equilibrium structure, then the local strain would be
$\sigma=d^2u_z/dx^2-1/R_{eq}=1/R-1/R_{eq}$. Equation~(\ref{eq1})
describes the corresponding local strain energy~\cite{Maceri2010}.
We find $k=7.2$~eV{\AA}$^2$ for the rigidity of the elastic beam
and $R_{eq}=24$~{\AA} for the optimum radius of curvature based on
PBE. The LDA values of $k=7.5$~eV{\AA}$^2$ and $R_{eq}=21$~{\AA}
are in fair agreement with the PBE values.

Same as in the infinite chain, the stability of the helical coil
is dominated by the covalent interatomic bonds, which are
described well by DFT calculations. The coil is further stabilized
by the weak attraction between neighboring strands that is similar
in nature to the inter-layer attraction in bulk black phosphorus.
As shown in superior Quantum Monte Carlo (QMC) calculations of the
latter system~\cite{DT250}, the fundamental nature of the
inter-layer interaction is rather non-trivial, different from a
van der Waals interaction, and not reproduced well by DFT
functionals with or without van der Waals corrections. When
compared to the more accurate QMC value of $81$~meV/atom, the LDA
value of $94$~meV/atom overestimates and the PBE value of
$16$~meV/atom underestimates the inter-layer interaction in bulk
black phosphorus. We also notice the large ratio of $5-6$ between
PBE and LDA values for the weak inter-layer interaction.
Extrapolating what is known about the interlayer interaction in
black phosphorus to the inter-chain interaction in a 2D assembly
of chains of Fig.~\ref{fig1}(d) or the related wall of the helical
coil in Fig.~\ref{fig1}(c), we expect that PBE will also
underestimate and LDA overestimate the value of this weak
interaction.

In view of the fact that the optimum coil radius is much larger
than the chain thickness, the wall of the infinite helical coil in
Fig.~\ref{fig1}(c) is well represented by a 2D assembly of chains
of Fig.~\ref{fig1}(d). We found that the most stable 2D
arrangement is one with AB stacking of chains. The inter-chain
interaction energy ${\Delta}E$ is displayed as a function of the
inter-chain distance $d$ in Fig.~\ref{fig2}(b). As anticipated
above, we expect a large difference between PBE and LDA
interaction energies. We obtain the optimum distance
$d_{eq}=7.3$~{\AA} and the interaction energy
${\Delta}E=5.3$~meV/atom based on PBE. LDA suggests a smaller
separation $d_{eq}=6.5$~{\AA} and a much larger interaction energy
${\Delta}E=30.8$~meV/atom. While still small, the LDA interaction
energy is roughly five times higher than the PBE value.

\begin{figure}[t!]
\includegraphics[width=1.0\columnwidth]{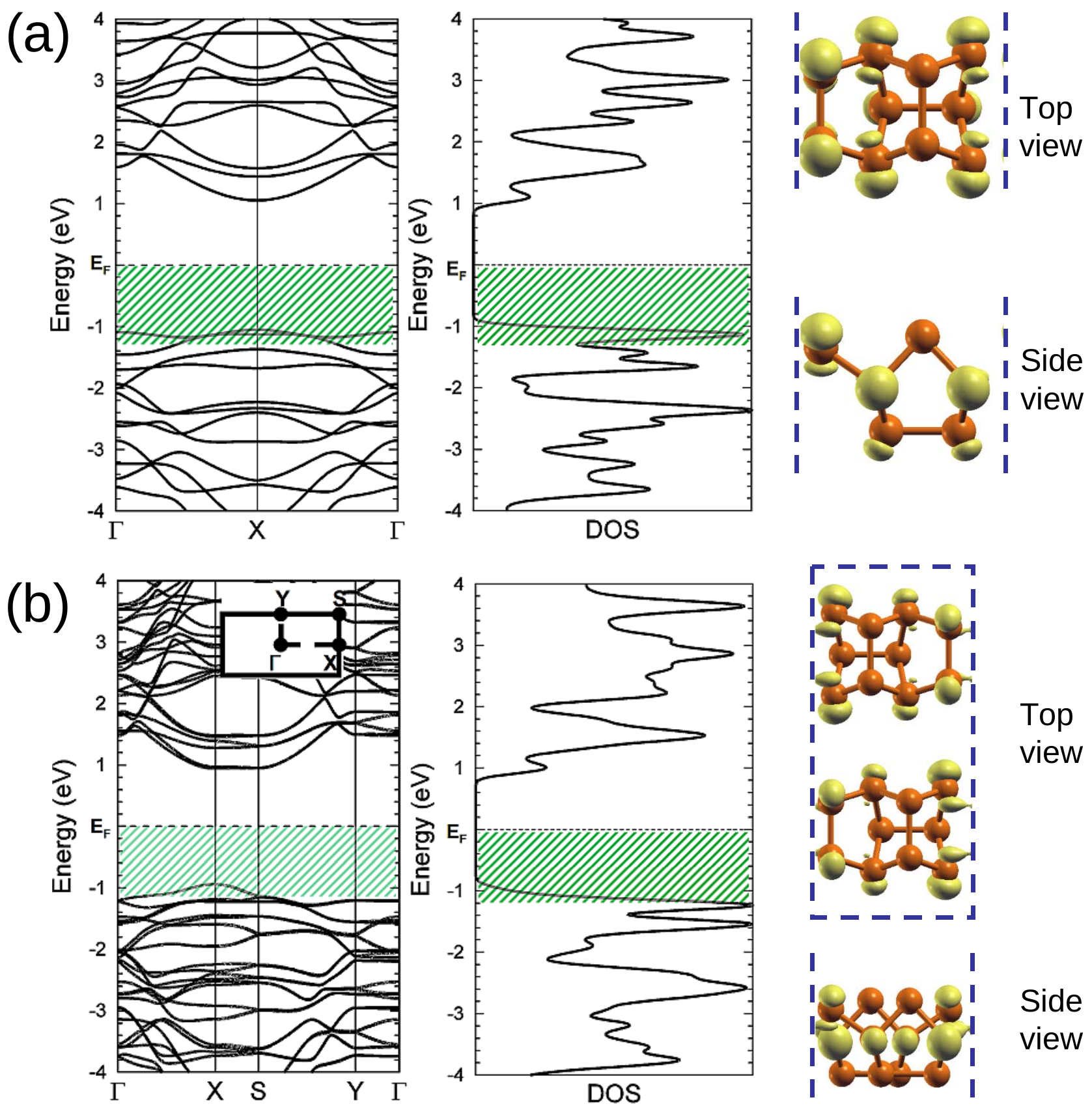}
\caption{(Color online) Electronic structure of (a) an isolated P
chain and (b) a 2D layer of P chains, shown in
Figs.~\protect\ref{fig1}(a) and \protect\ref{fig1}(d). Left panels
depict the electronic band structure based on PBE and the middle
panels the corresponding density of states. The Brillouin zone is
shown as inset of the left panel in (b). The right panels depict
the charge distribution associated with frontier states in the
valence band region, indicated by the green hashed region in (a)
and (b), which extends from $E_F$ to $0.2$~eV below the top of the
valence band. Charge density contours are superposed to structural
models, with the unit cells indicated by the dashed lines. Due to
differences in the density of states between these systems, the
contours are presented at the electron density $0.008$~e/{\AA}$^3$
in (a) and $0.002$~e/{\AA}$^3$ in (b). \label{fig3}}
\end{figure}

\section{Likely synthesis scenario}

Postulating a new allotrope is of limited use without a plausible
formation scenario. We note that previously, the void inside
carbon nanotubes has been successfully filled by sublimed C$_{60}$
fullerenes that eventually fused to an inner
nanotube~\cite{Bandow01}. Similarly, functionalized diamondoid
molecules were observed to enter the nanotube void, where they
converted to carbon chains~\cite{DT216} or diamond
nanowires~\cite{DT219}. Inspired by these results, we feel that
the most suitable scenario to form a helical coil phosphorus
allotrope involves a cylindrical cavity, shown in
Fig.~\ref{fig1}(e).

Suitable cavities with an optimum inner diameter of few nanometers
may be found in zeolites or in nanotubes of carbon, BN and other
materials. The phosphorus feedstock could be white, red or violet
phosphorus that had been sublimed in vacuum, under exclusion of
air. The sublimed species would likely be finite-chain segments,
shown in Fig.~\ref{fig1}(b), which may enter at the open end and
benefit energetically from the interaction with the inner wall of
the cavity. As seen in Fig.~\ref{fig2}(c), where we consider the
related system of an isolated chain on graphene, this interaction
is weak and similar in nature to the inter-chain interaction in
Fig.~\ref{fig2}(b). The optimum arrangement is found by inspecting
the adsorption energy ${\Delta}E$ as a function of height $h$ in
Fig.~\ref{fig2}(c). For the optimum geometry, we find
${\Delta}E_{eq}=9$~meV per P atom at $h_{eq}=3.9$~{\AA} based on
PBE and ${\Delta}E_{eq}=44$~meV per P atom at $h_{eq}=3.3$~{\AA}
based on LDA. We notice here again the adsorption energy ratio of
${\approx}5$ between LDA and PBE results, consistent with our
other results.

Once inside and near the wall of the cylindrical cavity, finite
P$_n$ chain fragments will benefit energetically from an
end-to-end connection that eliminates open ends. The number of
atoms in the finite circular arc, preferentially oriented along
the perimeter of the inner cavity, will grow. At the elevated
temperatures of subliming phosphorus, the growing P$_n$ ring
fragments are very unlikely to interconnect with corresponding
segments that contain exactly the right number of atoms, which
would complete a ring at the optimum distance to the wall. Much
more likely, the last segment to join before possible ring closure
will be too long and start the formation of a helical coil. Since
transformation of the coil to one or more adjacent rings would
require bond breakage within the coil, it is unlikely to happen.

\section{Electronic structure results}

The electronic structure of the new allotrope, similar to that of
phosphorene, is of utmost interest~\cite{DT229}. Our PBE results
for the related phosphorus chains and their 2D assemblies are
shown in Fig.~\ref{fig3}. As seen in Fig.~\ref{fig3}(a), the
P$_\infty$ chain has a direct fundamental band gap of $2.03$~eV at
$X$. Similarly, also the 2D chain assembly has a direct band gap
of $1.82$~eV at $X$ as seen in Fig.~\ref{fig3}(b). Based on what
is known theoretically and experimentally about few-layer
phosphorene~\cite{DT229}, the PBE band gap values are strongly
underestimated in comparison to the experiment.

In view of the fact that phosphorus is typically $p-$doped, we are
also interested in the nature of the frontier states in the
valence band region. We plotted the charge distribution of these
states, covering the energy range between $E_F$ and $0.2$~eV below
the top of the valence band, in the right panels of
Fig.~\ref{fig3}. Similar to what is known about black phosphorus,
we observe lone-pair electron states in the isolated chain in
Fig.~\ref{fig3}(a) that contributes to the electronic inter-chain
coupling modifying the band structure, as seen in
Fig.~\ref{fig3}(b). This demonstrated influence of the inter-chain
coupling on the electronic structure is a clear evidence that the
interaction differs from a purely van der Waals interaction,
similar to black phosphorus~\cite{DT250}.

\section{Discussion}

In view of the relatively low beam rigidity of the coiling chain,
we expect the coils to adjust their radius freely for the optimum
fit inside cylindrical cavities. The ability of the helical coil
strands to slide past each other allows the helix to adjust to a
changing cavity diameter. In view of the favorable inter-chain
interaction of $16$~meV/atom for an AB stacking in the radial
direction, with an optimum inter-chain distance of $0.6$~nm based
on PBE, we consider it quite possible for a second helix forming
inside the outer helix. Considering an outer helical coil at its
equilibrium radius $R_{out}=2.4$~nm, the inner coil should have a
radius of $R_{in}{\approx}1.8$~nm. In PBE, the strain energy in
the inner structure of $<1$~meV/atom according to
Fig.~\ref{fig2}(a) is negligibly small when compared to the
additional inter-chain interaction energy of $16$~meV/atom. We may
even imagine additional helices forming inside the double-helix
structure. In view of the low beam rigidity of the chain and the
large inter-chain interaction, a structure consisting of nested
coaxial coils should be even more stable than bulk black
phosphorus, the most stable phosphorus allotrope known to date.

As suggested by the end-on view of a chain in Fig.~\ref{fig1}(a),
the cross-section of the helical coil should appear as lines of
pentagons near the walls and along the axis of the cylindrical
cavity in Transmission Electron Microscopy (TEM) images. This is
very similar to recently observed TEM data~\cite{ZhangPNW}. We can
imagine left- and right-handed helical coils forming
simultaneously and coexisting inside a suitable cylindrical
cavity.

As a structural alternative, a black phosphorus monolayer may also
roll up to a tube inside a carbon nanotube with a $2.4$~nm radius.
Assuming an inter-wall distance of $0.5$~nm, the radius of the
phosphorene nanotube should be $R{\approx}1.9$~nm, and its strain
energy should be ${\approx}8.6$~meV/atom if bent along the soft
direction to become an armchair P nanotube, or $42.1$~meV/atom if
bent along the normal, harder direction~\cite{DT255}. Thus,
energetically, a black phosphorus nanotube is not favorable. Also
a straight 1D phosphorus chain inside a nanotube should be less
stable by $>6$~meV/atom than the coiled structure. Thus, we find
the coil to be more stable than competing phosphorene nanotube and
straight chain structures. Also, assuming that the new phase forms
by sublimation of red phosphorus, it will more likely resemble
structural elements of red P than the completely different black
P.

In layered black phosphorus, the observed bulk band gap value of
$0.35$~eV is known to increase to $2$~eV in the monolayer due to
the change in the weak interlayer interaction. PBE calculations
underestimate the band gap significantly, suggesting a value of
$0.04$~eV for the bulk and $0.9$~eV for the monolayer. We thus
expect also the calculated band gap in isolated chains to lie
below the experimental value and to decrease due to inter-chain
interaction in 2D chain assemblies. In helical coils, a further
reduction, which should furthermore depend on the coil radius, is
expected due to improved screening of the electron-hole
interaction. In view of this reasoning, it is not surprising that
our calculated band gap values lie rather close to the value of
$1.95$~eV that has been observed in fibrous red
phosphorus~\cite{Fasol85}.

Further electronic structure changes induced by coil deformation
may open a wide range of applications. Similar to bulk black
phosphorus, where changes in the interlayer distance $d$ modify
the band gap $E_g$ significantly, we find that axial compression
of the coil also modifies the band gap. At the optimum value
$0.73$~nm for the inter-coil distance $d$, defined in
Fig.~\ref{fig2}(b), the gap is direct and $E_g=1.8$~eV according
to Fig.~\ref{fig3}(b). Reducing $d$ to $0.6$~nm turns the gap
indirect and reduces its value to ${\approx}0.8$~eV. Further
reduction to $d=0.5$~nm turns the coil metallic. Increasing the
inter-chain distance to $d=0.8$~nm opens the gap to $E_g~2.0$~eV,
while maintaining its direct character. Of course, these changes
in the fundamental band gap may only be exploited inside
semiconducting nanotubes, such as BN, which have even larger band
gaps.

\section{Conclusions}

In conclusion, we have identified computationally an unusually
stable helical coil allotrope of phosphorus. Our {\em ab initio}
Density Functional Theory calculations indicate that the uncoiled,
isolated straight 1D chain is equally stable as a monolayer of
black phosphorus dubbed phosphorene. The coiling tendency and the
attraction between adjacent coil segments add an extra
stabilization energy of ${\approx}12$~meV/atom to the coil
allotrope, similar in value to the ${\approx}16$~meV/atom
inter-layer attraction in bulk black phosphorus. Thus, the helical
coil structure is essentially as stable as black phosphorus. In
view of the low beam rigidity of the chain and the large
inter-chain interaction, a structure consisting of nested coaxial
coils should be even more stable than bulk black phosphorus, the
most stable phosphorus allotrope known to date. With an optimum
radius of 2.4~nm, the helical coil of phosphorus may fit well and
even form inside wide carbon nanotubes. We find the coiled P
structure to be a semiconductor with a direct gap exceeding
$1.8$~eV. The size and character of the band gap can further be
modified by small structural changes in the coil.

\section{Methods}

Global search for small structural fragments of phosphorus was
performed using Adaptive Genetic Algorithms based on a
tight-binding Hamiltonian with universal
parameters.~\cite{DLiuTB16} Suitable structural candidates were
then optimized using {\em ab initio} density functional theory
(DFT) as implemented in the {\textsc SIESTA}~\cite{SIESTA} code to
obtain insight into the equilibrium structure, stability and
electronic properties of phosphorus structures reported in the
main manuscript. All isolated structures, including infinite 1D
chains and bent chain segments, have been represented using
periodic boundary conditions and separated by a $15$~{\AA} thick
vacuum region in all directions. We have used the
Perdew-Burke-Ernzerhof (PBE)~\cite{PBE} or alternately the Local
Density Approximation (LDA)~\cite{{Ceperley1980},{Perdew81}} forms
of the exchange-correlation functional, norm-conserving
Troullier-Martins pseudopotentials~\cite{Troullier91}, and a local
numerical double-$\zeta$ basis including polarization orbitals.
The Brillouin zone of periodic structures has been sampled by a
fine grid~\cite{Monkhorst-Pack76} of $12{\times}1{\times}1$
k-points for 1D structures and $12{\times}8{\times}1$ k-points for
2D structures. We found our basis, our $k-$point grid, and the
mesh cutoff energy of $180$~Ry used in the Fourier representation
of the self-consistent charge density to be fully converged,
providing us with a precision in total energy of $2$~meV/atom. All
geometries have been optimized using the conjugate gradient
method~\cite{CGmethod} until none of the residual Hellmann-Feynman
forces exceeded $10^{-2}$~eV/{\AA}.

\begin{suppinfo}
A movie depicting schematically the formation mechanism of a
helical coil of phosphorus inside a cylindrical cavity.\\
\end{suppinfo}
\quad\par

{\noindent\bf Author Information}\\

{\noindent\bf Corresponding Author}\\
$^*$E-mail: {\tt tomanek@pa.msu.edu}

{\noindent\bf Notes}\\
The authors declare no competing financial interest.

\begin{acknowledgement}
We acknowledge the assistance of Garrett B. King with the
graphical representation of evolving structures. This study was
supported by the NSF/AFOSR EFRI 2-DARE grant number
\#EFMA-1433459. Computational resources have been provided by the
Michigan State University High Performance Computing Center.
\end{acknowledgement}

%

\begin{mcitethebibliography}{35}
\bibitem[Bra()]{Brandt1669}
Elemental phosphorus was first extracted from human urine by
German alchemist
  Hennig Brand around 1669.\relax
\mciteBstWouldAddEndPunctfalse
\mciteSetBstMidEndSepPunct{\mcitedefaultmidpunct}
{}{\mcitedefaultseppunct}\relax \EndOfBibitem
\end{mcitethebibliography}

\providecommand{\latin}[1]{#1}
\providecommand*\mcitethebibliography{\thebibliography} \csname
@ifundefined\endcsname{endmcitethebibliography}
  {\let\endmcitethebibliography\endthebibliography}{}

\end{document}